\documentstyle[twocolumn,prl,aps]{revtex}

\begin{document}
\draft
\catcode`=11 \catcode`=12 
\twocolumn[\hsize\textwidth\columnwidth\hsize\csname@twocolumnfalse\endcsname
\title{The role of Berry phase in the spectrum of order parameter dynamics:\\
a new perspective on Haldane's conjecture on antiferromagnetic spin chains}
\author{Yue Yu$^1$ and Qian Niu$^2$}
\address{1. Institute of Theoretical Physics, Chinese Academy of Sciences, Beijing 100080, P. R. China}
\address{2. Department of Physics, University of Texas, Austin, Texas 78712-1081 }
\maketitle
\begin{abstract}        

We formulate the dynamics of local order parameters by extending the recently
developed adiabatic spinwave theory involving the Berry curvature, and derive a
formula showing explicitly the role of the Berry phase in determining the spectral
form of the low-lying collective modes. For antiferromagnetic spin chains,
the Berry phase becomes a topological invariant known as the Chern number.   
Our theory predicts the existence of the Haldane gap for a topologically trivial 
ground state, and a linear dispersion of low-lying excitations for a 
non-trivial ground state.  

\end{abstract}
\pacs{PACS numbers:75.30.Ds,75.50 Ee,75.30 Fv,75.40 Gb,75.10 Jm,03.65 Bz}]

Ever since Landau's formulation of continuous phase transition, the study of 
order parameter and its associated dynamics has occupied a central part of 
modern physics.  Dynamics of the order parameter gives rise to collective 
excitations known as Goldstone modes, with important consequences on thermal,  
mechanical, electrical or magnetic properties of physical systems.  Also, 
the absence of symmetry breaking in lower dimensional systems at finite 
temperatures can be understood as a result of thermal fluctuations of the 
low-lying modes of the order parameter dynamics.  For many one-dimensional
systems, there exist well-defined collective modes, such as the `spinwaves' 
in the antiferromagnetic Heisenberg spin chain \cite{DCP}, even though 
the ground state is disordered. It is very attempting to regard these 
collective modes also as that of the order parameter dynamics while the 
destruction of the long range ordering in the ground state is attributed to their 
quantum fluctuations \cite{cole}. 

In this Letter, we formulate a theory of order parameter dynamics based on the 
local ordering in the system alone.  We follow the approach of Ref.\cite{niu} 
for symmetry breaking magnetic systems to derive the equations of
motion and a formula for the collective excitation spectrum:
\begin{equation}
\hbar \omega =\frac{\Delta E}{B},  \label{hbo}
\end{equation}
where $\Delta E$ is the energy increase from the ground state 
for a frozen configuration of the order parameter and 
$B$ is the Berry phase of the many-body wave function
during a cycle of the collective motion.  We apply our theory to the systems of 
antiferromagnetic spin chains, showing that the presence or absence of a 
Haldane gap \cite{hald} is directly tied to a topological charge in the 
ground state. 

Haldane conjectured that the excitations of an antiferromagnetic  
Heisenberg chain have a gap for spins of integer $S$ and are gapless for 
spins of half-integer values \cite{hald}.  This was based on a mapping to a nonlinear
sigma-model in the large $S$ limit, where a topological action term is present 
for half-integer spins but not for integer spins.  Without the topological term,
the nonlinear sigma-model was known to be gapped, but a rather elaborate 
renormalization group analysis was needed to show that the presence of the 
topological term can render the excitations gapless \cite{frad}. Haldane's
conjecture seems to be correct also for small spins, because it conforms with the
exact solutions for the extreme cases of $S=1/2$ \cite{DCP} and $S=1$ \cite{aff} 
and with numerical results. The success of Haldane's conjecture is highly 
celebrated in the theoretical 
physics community, because it gives a prime example that topology 
can play such a decisive role in measurable effects.

Here we present a direct mechanism showing how topology works its way 
to determine the spectral form of the excitations.  
For the antiferromagnetic chains, we will show
that the Berry phase for a mode of wave number $k$ can be written for small $k$ as 
\begin{equation}
B\propto \frac {kL}{2\pi } Q+O(k^2),  \label{bk}
\end{equation} 
where $L$ is the length of the chain and $Q$ is the topological charge defined as 
the Chern number of the wave function mapped to the order parameter configuration
in the excitation mode.  On the other hand, the energy increase in a frozen 
configuration of the order parameter should have the form

\begin{equation}
\Delta E \propto L k^2.  \label{de}
\end{equation}
Therefore, depending on the presence or absence of this topological charge $Q$,
the excitation spectrum is linearly dispersed for small $k$ or becomes gapped: 
\begin{equation}
\hbar \omega \propto \cases{
k, & if $Q\ne 0$; \cr 
\Delta , & if $Q=0$,} \label{dis}
\end{equation}
where $\Delta $ is a constant. In many one dimensional antiferromagnetic
models of integer spin, such as the AKLT model for $S=1$  \cite{aff} 
and its SU$(N)$ generalization \cite{Read}, the exactly soluble ground states 
are topologically trivial ($Q=0$).  On the other hand, several spin half-integer
models have been constructed with topologically non-trivial ground states, e.g.,
the resonating-valence-bond ground state with a twofold degeneracy in the 
spin-Pereils order \cite{Read}. The Lieb-Schultz-Mattis theorem for 
spin-half and its generalization to arbitrary half-integer spins \cite{LSM} 
also indicated this non-triviality.  In lights of these facts and the
general arguments given in Ref.\cite{aff1}, we can thus conclude that our spectral 
formula is really consistent with Haldane's conjecture \cite{hald}.  

The standard procedure of introducing the order parameter is to apply a
weak external field to force the system to order in a particular way, corresponding 
to some non-zero expectation value of the operator in conjugation to the field
$\varphi^j_x=\langle\hat O^j_ x\rangle$, where $x$ denotes the position and $j$
labels the internal components.  If the ordering persists after the field is
turned-off, we say that there is a spontaneous symmetry breaking, and the nonzero 
expectation value is called the order parameter. Standard examples of order parameter 
include the magnetization field in magnetic materials, the condensate wave 
function for superfluids \cite{fey}, the Ginzburg-Landau order parameters in
superconductivity \cite{gl} and many other condensed matter systems
\cite{and}. To facilitate the discussion of its 
dynamics, we generalize the notion by defining the order parameter in any state 
simply as the expectation value of $\hat O^j_ x$ in that state.  In this way, 
we can also talk about the order parameter even for systems without long range order.  

As long as there is a strong local ordering, the low-lying excitations should be 
dominated by the order parameter dynamics in the following sense.  Consider the set 
of constrained ground states defined as the union of the lowest energy state for each 
configuration of the order parameter.  If an initial state prepared from this set 
will evolve entirely within this set, then we have a closed dynamics of the order
parameter because such states are labeled uniquely by the order parameter 
configuration.  We assume this is the case, which can be justified at least for those 
long wavelength deviations of the ground state configuration.  
We can then apply the time dependent variational principle along the line of 
Refs.\cite{niu,niu2} to derive the equations of motion  of the order parameter
dynamics
\begin{equation}
\sum_{j^{\prime}, x^{\prime}}\hbar\Omega_{ x x
^{\prime}}^{jj^{\prime}} \dot\varphi^{j^{\prime}}_{ x^{\prime}} = \frac{
\partial E}{\partial \varphi^j_x },  \label{el}
\end{equation}
which involve the energy $E=\langle\psi|H|\psi\rangle$ of the constrained ground state
and the Berry curvature 
\begin{equation}
\Omega_{ x x^{\prime}}^{jj^{\prime}}=\frac{\partial}{\partial\varphi^j_x}
\langle\psi|\frac{i\partial} {\partial \varphi^{j^{\prime}}_{ x^{\prime}}}
|\psi\rangle-\frac{\partial}{\partial \varphi^{j^{\prime}}_{ x^{\prime}}} \langle\psi|
\frac{i\partial}{\partial \varphi_{ x}^{j}}|\psi\rangle.  \label{bc}
\end{equation}

The spectral formula (\ref{hbo}) can be derived directly from the 
equations of motion. In Ref. \cite{niu2}, the formula was obtained for 
the case of spinwave by linearizing the equations of motion around the ground state 
for ferro-, ferri- and antiferromagets.  There it was also shown that this 
Berry phase is actually given by the reduction of the total magnetization from the 
ground state due to the spinwave,  thus proving and generalizing an earlier 
result from Ref. \cite{niu} for the spinwave  spectrum.  This formula has 
now served the basis for a number of successful first principle calculations 
for ferromagnetic crystals \cite{calc}, and similar work on other types of magnetic 
materials are expected in the near future.  Exactly the same derivation of the 
spectral formula can be applied for the order parameter dynamics of any system 
with a symmetry breaking ground state.  The same reasoning should also give 
the Berry phase in terms of the deviation from the ground
state expectation value of the generator of the collective motion.

For a system with no spontaneous breaking of symmetry, such as the antiferromagnetic 
chain, the spectral formula (\ref{hbo}) still stands 
as shown by the following arguments. We multiply both side 
of (\ref{el}) by $dt\, \delta_E \varphi_{ x}^j$ and sum over $x$ and $j$, i.e.
\begin{equation}
\sum_{a,a'}\hbar 
\Omega _{ a a'} \frac{\partial \varphi^{a'}}{\partial t} dt\,\delta_E \varphi^a
=\sum_a\frac{ \partial 
E}{\partial \varphi^a}\, dt\, \delta_E \varphi^a,  \label{times}
\end{equation}
where the two labels have been condensed into one for simplicity 
($\varphi^a\equiv \varphi^j_x$), and $\delta_E \varphi^a$ is the variation 
in a direction perpendicular to the constant energy trajectory of the order 
parameter.  We then integrate (\ref{times}) over the two-dimensional domain 
${\cal D}_\varphi$ consisting of a one parameter family of trajectories 
ranging from the fixed point in the absolute ground state to a trajectory 
${\cal C}_\varphi$ of 
finite amplitude of the collective motion.  In the harmonic regime, where 
the collective modes may be defined, we expect that the time period $T$ 
to be a constant, so that the integration yields
\begin{equation}
\hbar \int_{{\cal D}_\varphi}\sum_{aa'}\delta_t \varphi^a
\delta_E \varphi^{a'}\Omega _{aa'}=T\Delta E, \label{T}
\end{equation}
where $\Delta E=E-E_0$ is the energy increase from the ground state, and 
$\delta_t \varphi^a$ denotes the
variation of the order parameter along the trajectory (i.e.,
$\delta_t=dt\partial/\partial t$).
Because the time period $T$ relates to the frequency $\omega$ of the
collective mode by $T=\frac{2\pi}{\omega}$,  we  
arrive at the formula (\ref{hbo}) with the Berry phase given by 
\begin{eqnarray}
B&=&\frac{1}{2\pi}\int_{{\cal D}_\varphi}\sum_{aa'}\delta_t
\varphi^a\delta _E \varphi^{a'}\Omega _{aa'} \nonumber\\
&=&\frac{1}{2\pi}\oint_{{\cal C}_\varphi}\sum_a
\delta_t \varphi^a\langle\psi|\frac{i\partial}
{\partial \varphi^a}|\psi\rangle.
\label{berry}
\end{eqnarray}
where the second equality results from the Stokes theorem.

To appreciate how the Berry phase determines the spectral form of the 
low-lying collective mode, we expand it in powers of the 
wave number $k$
\begin{equation}
B=B_0+B_1k+B_2k^2+... .  \label{bex}
\end{equation}
For ferro- and ferrimagnets, we have $B_0\ne 0$ because the total magnetization 
reduction due to a spinwave is nonzero even in the limit of zero $k$.  
This yields a quadratic dispersion for the spectrum at small $k$ in light of 
Eq.(3).  For lattices with antiferromagnetic ordering, we have $B_0=0$ due 
to sublattice symmetry \cite{Read}, while $B_1\ne 0$ \cite{Kittel} from a careful 
analysis of magnetization reduction in the presence of a spinwave of small but 
nonzero $k$.  This reproduces the standard result that antiferromagnetic 
spinwaves have a linear  dispersion at small $k$.  

For antiferromagnetic spin chains, where the spin rotation symmetry cannot be
spontaneously broken according to Coleman's theorem, the results drawing from the 
total spin reduction may not be applicable.  Fortunately, we have two 
observations that help to establish the topological interpretation of 
the Berry phase shown in (2). First, the total Berry phase
 due to a cyclic motion in the order parameter 
configuration space can be expressed as a sum of the Berry phases due to the cyclic 
motion of local order parameter at each site.  In other words, we may write 
the Berry phase (\ref{berry}) in the form
\begin{equation}
B=\sum_x \frac {1}{2\pi} \oint_{ C_x} 
\delta_t \vec \varphi_x\cdot\langle \psi|\frac{i\partial}
{\partial \vec \varphi_x}|\psi\rangle,
\label{berry2}
\end{equation}
where $C_x$ denotes the path of the local spin moment $\vec \varphi_x$ 
as the projection of the configurational path ${\cal C}_\varphi$ on site $x$.
For each term in the sum, the constrained ground state $|\psi\rangle$ is evaluated 
with the spin moments on all sites except $x$ set to their true ground state 
value, i.e., zero.  This observation may fail for itinerant spin systems such 
as the $t-J$ model, because the Berry curvature is known to have inter-site 
terms which prevent the resolution of the Berry phase into contributions 
from each site.  However, we expect the observation to be true for localized 
spin systems such as the Heisenberg model.  

A direct consequence of the above observation is that the total Berry phase
(\ref{berry2}) can be written as the number of the space
periods, $n={kL\over 2\pi}$, times the Berry phase in one period, \cite{note} 
\begin{equation}
B=\frac k{2\pi }L B_{{\rm p}},  \label{berry3}
\end{equation}
where $L$ is the size of the chain and $B_{p}$ is defined by (\ref{berry2})
but with the sum over $x$ confined in one space period $(0,\lambda)$.
Our second observation then shows that $B_p$ is proportional to a topological charge. 
We note that because of the local antiferromagnetic ordering, the directions of 
spin moments on neighboring sites tend to be opposite to each other, so is 
the sense of chirality of their motion. Therefore, 
the contributions to the Berry phase from neighboring
sites almost cancel each other for long wavelength collective modes.  
It will thus be convenient to introduce the staggered order parameter 
$\vec m_x=(-1)^x \vec \varphi_x$, so that the Berry phase per period becomes
\begin{equation}
B_p=\frac{1}{2\pi}\sum_{x\in (0,\lambda)}  (-1)^x \oint _{C'_x} 
\delta_t\vec m_x\cdot \langle\psi|\frac{i\partial}{\partial \vec m_x}|\psi\rangle, 
\label{bp2}
\end{equation}
where $C'_x$ denotes the orbit of $\vec m_x$.  For small $k$, 
we may take the continuum limit 
by replacing the difference by a differential, 
\begin{eqnarray}
 &&B_p=\frac{1}{2\pi}\sum_{x\in (0,\lambda)} 
\oint_{ C'_x} \delta_t\vec m_x\cdot \sum_j \delta_x m^j_x
\nonumber\\
&&\biggl[\frac{\partial}{\partial m^j_x}\langle\psi|\frac{i\partial}{\partial\vec
 m_x}|\psi\rangle-\frac{\partial}{\partial \vec m_x}
\langle\psi|\frac{i\partial}{\partial m^j_x}|\psi\rangle\biggr], \label{bp3}
\end{eqnarray}
where $\delta_x$ stands for $dx {\partial\over \partial x}$ and the second term is an added zero term.   
Due to the spatial periodicity, the sum over $x$ corresponds to a closed loop integral,
so that (\ref{bp3}) becomes an integral over the closed space-time torus $T^2$
\begin{equation}
B_p\equiv Q=\frac{1}{2\pi}\oint_{T^2} \sum_{j,j'}\delta_x m^j_x \delta_t m^{j'}_x
\Omega_{jj'}(\vec m_x),  \label{bp4}
\end{equation}
with the curvature 
\begin{equation}
\Omega_{jj'}(\vec m)=\frac{\partial}{\partial m^j}\langle\psi 
|\frac{i\partial }{\partial
m^{j^{\prime}}}|\psi\rangle -\frac{\partial}{\partial
m^{j^{\prime}}}\langle\psi|\frac{i\partial }{\partial m^{j}}|\psi \rangle.
\end{equation}

Thus, the Berry phase per spatial period of a collective excitation in an antiferromagnetic  spin chain is in fact a topological invariant, the first Chern class 
for the mapping of the constrained ground state to the space-time structure of the 
local order parameter.  This reduces to the standard semiclassical result of 
\begin{equation}
Q=\frac{1}{2\pi}\oint_{T^2}  \delta_x \vec m \times \delta_t \vec m \cdot \vec m/|\vec m|^2,
\end{equation}
if we take $\Omega_{jj'}=\sum_l\epsilon_{jj'l}m^l/|\vec m|^2$.
Our expression (\ref{bp4}) is a generic result and model-independent. 
It has been expected that the topology of the ground states of spin chains is 
trivial for integer $S$ but non-trivial for half-integer $S$ \cite{aff1}. 
For an $S=1$ chain, the authors of
Ref.\cite{aff} provided an exact valence-bond-solid
ground state which is topologically trivial \cite{aff}. Read and Sachdev discussed
the SU($N$) antiferromagnetic chains in the large $N$ limit \cite{Read}. By using a trial ground
state wave function, they showed that there is a spin-Peierls order parameter
proportional to the topological charge of the ground state. They explicitly gave the
dependence of the ground state energies on the topology of the state. They concluded
that the valence-bond-solid ground state of the integer spin chains is topologically
trivial and not degenerate with a vanishing spin-Pereils order parameter; 
The resonating-valence-bond
ground state for the half-integer chains, on the other hand, 
is topologically nontrivial and degenerate due
to the different spin-Pereils 
order parameters. We can also see the topological property of the chains in the
twist introduced in \cite{LSM}, which, in some sense, gave a finite $S$ 
version of the topological term in the nonlinear sigma model \cite{aff2}. 

Our spectral formula eq.(\ref{hbo}) can also serve the basis for numerical calculation 
of the excitation spectrum.  
A quantitative comparison of the numerical result with known theoretical 
and experimental results should constitute a stringent test of our theory. For
example, in the spin-1/2 antiferromagnetic chain, the spinwave speed is $\pi/2$ times
larger than the semiclassical result \cite{DCP}, and it would be interesting to see
if the numerical calculation based on our formula will give the correct result.

In conclusion, we have formulated a theory of local order parameter dynamics 
and derived a formula for the spectral form of collective excitations 
in terms of the Berry phase.  For antiferromagnetic spin chains, we have shown 
in a  model-independent manner that the presence or absence of a gap is directly tied 
to the topological structure of the constrained ground state.  
For all known exact or model solutions of the spin chains, our result is consistent
with Haldane's conjecture. We also recognize that the topological consideration may
not be valid for itinerant spin systems because of the non-vanishing inter-site Berry
curvature.

The authors thank Wu-Ming Liu for the suggestion of this collaboration and
discussions. They are grateful to Ping Ao, Hong Chen, Sui-Tat Chui, E. Fradkin
and especially, Shao-Jing Qin and Zhao-Bin Su for useful discussions. 
QN thanks Institute of Theoretical Physics (Beijing) for the warm
hospitality, where the work was initiated. This work was supported
in part by the NSF (DMR 9614040, PHY 9722610) and NSF of China.


\begin{references}
\bibitem{note} This is also correct to the ferro- and ferrimagents. However,
due to summation or non-cancellation of the local order parameters of nearest neighbor
sites, $B_p$ is proportional to the wave length. Thus, the total Berry phase remains
a $k$-independent constant at the zeroth order.
\end{references}
\end{document}